\documentclass[journal,12pt,draftclsnofoot,onecolumn,twoside]{IEEEtran}

\usepackage{ifpdf}
 \ifpdf
     \usepackage[pdftex]{graphicx}
 \else
     \usepackage[dvips]{graphicx}
 \fi

\graphicspath{{figures/}}

\usepackage{cite}
\usepackage[cmex10]{amsmath}
\usepackage{amssymb,amsfonts}
\usepackage{algorithm}
\usepackage{algpseudocode}

\usepackage{mdwmath}
\usepackage{mdwtab}
\usepackage[belowskip=-15pt,aboveskip=1pt,small]{caption}
\IEEEaftertitletext{\vspace{-2.5\baselineskip}}
\begin{document}
%
\title{A Novel Algorithm for Rate/Power Allocation in OFDM-based Cognitive Radio Systems with Statistical Interference Constraints}

%
%
%
\author{{Ebrahim Bedeer, 
Octavia A. Dobre, 
Mohamed H. Ahmed, and
Kareem E. Baddour \IEEEauthorrefmark{2}}\\
\IEEEauthorblockA{Faculty of Engineering and Applied Science, Memorial University of Newfoundland,
St. John's, NL, Canada\\
\IEEEauthorrefmark{2} Communications Research Centre, Ottawa, ON, Canada\\
Email: \{e.bedeer, odobre, mhahmed\}@mun.ca, kareem.baddour@crc.ca}
}
\maketitle

\begin{abstract}
In this paper, we adopt a multiobjective optimization approach to jointly optimize the rate and power in OFDM-based cognitive radio (CR) systems. We propose a novel algorithm that jointly maximizes the OFDM-based CR system throughput and minimizes its transmit power, while guaranteeing a target bit error rate per subcarrier and a total transmit power threshold for the secondary user (SU), and restricting both co-channel and adjacent channel interferences to existing primary users (PUs) in a statistical manner. Since the interference constraints are met statistically, the SU transmitter does not require perfect channel-state-information (CSI) feedback from the PUs receivers. Closed-form expressions are derived for bit and power allocations per subcarrier. Simulation results illustrate the performance of the proposed algorithm and compare it to the case of perfect CSI.   Further, the results show that the performance of the proposed algorithm approaches that of an exhaustive search for the discrete global optimal allocations with significantly reduced computational complexity.
\end{abstract}

\begin{IEEEkeywords}
Bit and power allocation, cognitive radio, dynamic spectrum sharing, statistical interference constraints.
\end{IEEEkeywords}

%

\vspace*{-5pt}
\section{Introduction}
\vspace*{-3pt}
Cognitive radio (CR) can considerably enhance the spectrum utilization efficiency by dynamically sharing the spectrum between licensed/primary users (PUs) and unlicensed/secondary users (SUs) \cite{hossain2007cognitive}.
This is achieved by granting the SUs opportunistic access to the white spaces within the PUs spectrum, while controlling the interference to the PUs. Orthogonal frequency division multiplexing (OFDM) is recognized as an attractive modulation technique for CR due to its flexibility, adaptivity in allocating vacant radio resources, and spectrum shaping capabilities \cite{hossain2007cognitive}. A common technique to improve the performance of the OFDM-based systems is to dynamically load different bits and/or powers per each subcarrier according to the wireless channel quality and the imposed PUs interference constraints \cite{zhang2010efficient, bansal2008optimal, kang2009optimal, zhao2010power,  hasan2009energy, bansal2011adaptive}.



The prior work in the literature focused on maximizing the OFDM SU capacity/throughput while limiting the interference introduced to PUs to a predefined threshold \cite{zhang2010efficient, bansal2008optimal, kang2009optimal, zhao2010power,  hasan2009energy, bansal2011adaptive}.
The authors in \cite{bansal2008optimal, kang2009optimal, zhang2010efficient, bedeer2012adaptiveRWS} consider perfect channel-state-information (CSI) between the SU transmitter and the PUs receivers, which is a challenging assumption for practical scenarios.
In \cite{zhao2010power, hasan2009energy}, the authors assume only knowledge of the path loss for these links; however, such an assumption will cause the proposed algorithms to violate the interference constraints uncontrollably when applied in practice (i.e., since neither the instantaneous channel gains nor the channel statistics are known, there is no guarantee regarding the probability of violation of the interference constraints). The authors in \cite{bansal2011adaptive} assume knowledge of the channel statistics (i.e., the fading distribution and its parameters), which is a reasonable assumption for certain wireless environments, e.g., in non-line-of-sight urban environments, a Rayleigh distribution is usually assumed for the magnitude of the fading channel coefficients. 
In this paper, we adopt the same channel assumption as in \cite{bansal2011adaptive}; our main contributions  when compared with the work in the literature are as follows: 1) a multiobjective optimization approach\footnote{In a non-CR environment, jointly maximizing the throughput and minimizing the transmit power provides a significant performance improvement, in terms of the achieved throughput and transmit power, when compared to other work in the literature that separately maximizes the throughput (while constraining the transmit power) or minimizes the transmit power (while constraining the throughput), respectively \cite{bedeer2013joint}.} is used for the dynamic spectrum sharing problem and 2) we guarantee a certain OFDM SU bit error rate (BER).

That being said, in this paper we propose a novel low-complexity algorithm for OFDM-based CR systems that jointly maximizes the OFDM SU throughput and minimizes its transmit power, subject to a target BER per subcarrier and  total transmit power threshold for the SU, as well as statistical constraints on the  co-channel interference (CCI) and adjacent channel interference (ACI) to the PUs.
Closed-form expressions are derived for the bit and power allocations per subcarrier. Simulation results identify the performance degradation due to the incomplete channel information, by comparing the performance of the proposed algorithm with that of perfect CSI. Additionally, the results indicate that the performance of the proposed algorithm approaches that of an exhaustive search for the optimal allocations.

The remainder of the paper is organized as follows. Section II presents the system model and Section III introduces the proposed joint bit and power loading algorithm. Simulation results are presented in Section IV, while conclusions are drawn in Section V.

Throughout this paper we use bold-faced lower case letters for vectors, e.g., $\mathbf{x}$, and light-faced letters for scalar quantities, e.g., $x$. $[.]^T$ denotes the transpose operation, $\nabla$ represents the gradient operator, \textup{Pr}(.) denotes the probability, $\mathbb{E}[.]$ is the statistical expectation operator, $[x,y]^-$ represents $\textup{min}(x,y)$, and $\bar{\bar{\mathbb{X}}}$ is the cardinality of the set $\mathbb{X}$.

\section{System Model}
\vspace*{-2pt}
The available spectrum is assumed to be divided into $\mathcal{M}$ subchannels that are licensed to $\mathcal{M}$ PUs. A subchannel $m$, of bandwidth $B_m$, has $N_m$ subcarriers and $i_m$ denotes subcarrier $i$ in subchannel $m$, $i_m = 1, ..., N_m$. A PU does not occupy its licensed spectrum all the time and/or at all its coverage locations; hence, a SU may access such voids as long as no harmful interference occurs to adjacent PUs due to ACI, or to other PUs operating in the same frequency band at distant locations due to CCI.

\begin{figure}[t]
\centering
\includegraphics[width=0.50\textwidth]{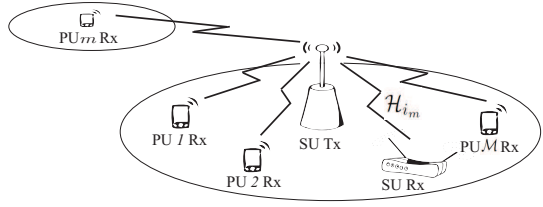}
\caption{Co-existence of an SU and $\mathcal{M}$ PUs in the spatial domain.}
\label{fig:sys}
\end{figure}

A typical CR system is shown in Fig.~\ref{fig:sys}. The SU first obtains the surrounding PUs information, such as the PUs positions and spectral band occupancies\footnote{This is done by visiting a database administrated by a government or third party, or by optionally sensing and determining the PUs radio frequency and positions, respectively \cite{zhao2010power}.}. Then, it makes a decision on the possible transmission subchannels. We consider that the SU has all the required information on the existing $\mathcal{M}$ PUs, and it decides to use the vacant $m$th PU   subchannel, $m\in \{1, ..., \mathcal{M}\}$.

Following the common practice in the literature, we assume that the instantaneous channel gains between the SU transmitter and receiver pairs are available through a delay- and error-free feedback channel \cite{zhang2010efficient, bansal2008optimal, kang2009optimal, zhao2010power,  hasan2009energy, bansal2011adaptive}. Additionally, we assume that the SU transmitter has knowledge of the fading distribution type and its corresponding parameters of the channels $\mathcal{H}_{sp}^{(\ell)}$ and $\mathcal{H}_{sp}^{(m)}$ to the $\ell$th and $m$th  PUs receivers, respectively (given the fact that estimating the instantaneous channel gains between the SU transmitter and the PUs receivers is practically challenging).

The interference, $\mathcal{J}_{i_m}$, from all the PUs to subcarrier $i_m$ of the SU is considered as in \cite{hasan2009energy, zhao2010power, bansal2008optimal, bansal2011adaptive}, and depends on the SU receiver windowing function and power spectral density of the PUs. On the other hand, the ACI depends on the power allocated to each SU subcarrier and the spectral distance between the SU subcarriers and the PUs. The ACI from the SU to the $\ell$th PU  receiver is formulated as \cite{hasan2009energy, zhao2010power, bansal2008optimal, bansal2011adaptive}
\begin{IEEEeqnarray}{c}
|\mathcal{H}_{sp}^{(\ell)}|^2 \sum_{i_m = 1}^{N_m} \mathcal{P}_{i_m} \varpi_{i_m}^{(\ell)} \leq \mathcal{P}_{\textup{ACI}}^{(\ell)},
\end{IEEEeqnarray}
where $\varpi_{i_m}^{(\ell)} = T_{s,m}  \: 10^{-0.1 L(d_{\ell})} \: \int_{f_{i_m,\ell}-\frac{B_\ell}{2}}^{f_{i_m,\ell}+\frac{B_\ell}{2}} \textup{sinc}^2(T_{s,m} f) \: df$, $T_{s,m}$ is the duration of the OFDM symbol of the SU, $L(d_{\ell})$ is the path loss in dB at distance $d_\ell$, $d_\ell$ is the distance between the SU and the $\ell$th PU  receiver, $f_{i_m,\ell}$ is the spectral distance between the SU subcarrier $i_m$ and the $\ell$th PU receiver frequency band,  $B_\ell$ is the bandwidth of the $\ell$th PU receiver, $\mathcal{P}_{i_m}$ is the transmit power per subcarrier $i_m$, $\mathcal{P}_{\textup{ACI}}^{(\ell)}$ is the interference threshold at the $\ell$th PU  receiver, and $\textup{sinc}(x) = \frac{\textup{sin}(\pi x)}{\pi x}$.

The CCI at the location of the distant $m$th PU receiver is required to be limited as

\begin{eqnarray}
|\mathcal{H}_{sp}^{(m)}|^2 \: 10^{-0.1L(d_m)} \sum_{i_m=1}^{N_m}  \mathcal{P}_{i_m} \leq   \mathcal{P}_{\textup{CCI}}^{(m)}, \label{eq:CCI_1}
\end{eqnarray}
where $\mathcal{P}_{\textup{CCI}}^{(m)}$ is the interference threshold at the $m$th PU. To reflect the SU transmitter's power amplifier limitations or/and to satisfy regulatory maximum power limits, the total SU transmit power is limited to a certain threshold $\mathcal{P}_{th}$ as \vspace*{-1pt}
\begin{eqnarray}
\sum_{i_m=1}^{N_m} \mathcal{P}_{i_m}\leq \mathcal{P}_{th}. \label{eq:CCI_2}
\end{eqnarray}

\vspace*{-5pt}

\section{Proposed Algorithm}
\subsection{Optimization Problem Formulation}
We propose a novel close-to-optimal algorithm that jointly maximizes the OFDM SU throughput and minimizes its transmit power, while satisfying a target BER per subcarrier\footnote{The constraint on the BER per subcarrier is a suitable formulation that results in similar BER characteristics when compared with an average BER constraint, especially at high signal-to-noise ratios (SNR) \cite{willink1997optimization}. Furthermore, it facilitates derivation of closed-form expressions for the optimal bit and power solutions, which reduces the algorithmic complexity.} and a total transmit power threshold for the SU, and limiting the introduced CCI and ACI to the $m$th and $\ell$th PUs  receivers below the thresholds $\mathcal{P}_{\textup{CCI}}^{(m)}$ and $\mathcal{P}_{\textup{CCI}}^{(\ell)}$ with at least a probability of $\Psi_{\textup{CCI}}^{(m)}$ and $\Psi_{\textup{ACI}}^{(\ell)}$, respectively. 
The optimization problem is formulated as
\begin{IEEEeqnarray}{c}
\underset{\mathcal{P}_{i_m}}{\textup{Minimize}} \: \sum_{{i_m} = 1}^{N_m}\mathcal{P}_{i_m} \:\: \textup{and} \:\:  \underset{b_{i_m}}{\textup{Maximize}} \: \sum_{{i_m} = 1}^{N_m}b_{i_m}, \nonumber 
\end{IEEEeqnarray}
{\small{
\begin{subequations}
\begin{IEEEeqnarray}{rcl}
\hspace{1.0cm}\textup{subject to} \hfill \textup{BER}_{i_m} &{} \leq {}& \textup{BER}_{th,{i_m}},  \IEEEeqnarraynumspace \label{eq:eq_first} \\
\sum_{i_m=1}^{N_m} \mathcal{P}_{i_m} &{} \leq {}& \mathcal{P}_{th},  \label{eq:eq_1} \\
\textup{Pr}\left(|\mathcal{H}_{sp}^{(m)}|^2 10^{-0.1L(d_m)} \sum_{i_m=1}^{N_m}  \mathcal{P}_{i_m}   \leq  \mathcal{P}_{\textup{CCI}}^{(m)}\right) & \geq& \Psi_{\textup{CCI}}^{(m)},   \label{eq:eq_1_CCI} \\
\textup{Pr}\left(|\mathcal{H}_{sp}^{(\ell)}|^2 \sum_{i_m = 1}^{N_m}  \mathcal{P}_{i_m} \varpi_{i_m}^{(\ell)} \leq  \mathcal{P}_{\textup{CCI}}^{(\ell)}\right) & \geq & \Psi_{\textup{ACI}}^{(\ell)},  \label{eq:eq_1_2}
\end{IEEEeqnarray}
\label{eq:op_1_basic}
\end{subequations}
}}\hspace{-4pt}where $i_m = 1, ..., N_m,  \ell = 1, ..., \mathcal{M}$, $b_{i_m}$ is the number of bits per subcarrier $i_m$, and $\textup{BER}_{i_m}$ and $\textup{BER}_{th,{i_m}}$  are the BER per subcarrier $i_m$ and the threshold value of the BER per subcarrier $i_m$, respectively.

A non-line-of-sight propagation environment is assumed; therefore, the channel gains $\mathcal{H}_{sp}^{(m)}$ and $\mathcal{H}_{sp}^{(\ell)}$ can be modeled as zero-mean complex Gaussian random variables, and, hence, $|\mathcal{H}_{sp}^{(m)}|^2$ and $|\mathcal{H}_{sp}^{(\ell)}|^2$ follow an exponential distribution \cite{bansal2011adaptive}. Accordingly, the statistical CCI interference constraint in (\ref{eq:eq_1_CCI}) can be evaluated as
\begin{IEEEeqnarray}{RCL}
1 - \exp\left(- \frac{\nu}{10^{-0.1L(d_m)} \sum_{i_m=1}^{N_m} \mathcal{P}_{i_m}}\mathcal{P}_{\textup{CCI}}^{(m)}\right) \geq \Psi_{\textup{CCI}}^{(m)}, \label{eq:stat_1}
\end{IEEEeqnarray}
where $\frac{1}{\nu}$ is the mean of the exponential distribution. Eqn. (\ref{eq:stat_1})  can be further written as
\begin{IEEEeqnarray}{RCL}
\sum_{i_m=1}^{N_m}  \mathcal{P}_{i_m} \leq   \frac{\nu 10^{0.1L(d_m)}}{-\ln(1 - \Psi_{\textup{CCI}}^{(m)})} \mathcal{P}_{\textup{CCI}}^{(m)}, \label{eq:stat_2}
\end{IEEEeqnarray}
and the constraints in (\ref{eq:eq_1}) and (\ref{eq:eq_1_CCI}) can be combined as

\begin{IEEEeqnarray}{RCL}
\sum_{i_m=1}^{N_m}  \mathcal{P}_{i_m} \leq \Big[\mathcal{P}_{th},  \frac{\nu 10^{0.1L(d_m)}}{-\ln(1 - \Psi_{\textup{CCI}}^{(m)})} \mathcal{P}_{\textup{CCI}}^{(m)} \Big]^-.
\end{IEEEeqnarray}
Similarly, the statistical ACI constraint in (\ref{eq:eq_1_2}) is rewritten as
\begin{IEEEeqnarray}{RCL}
\sum_{i_m=1}^{N_m}  \mathcal{P}_{i_m} \varpi_{i_m}^{(\ell)} \leq  \frac{\nu}{-\ln(1 - \Psi_{\textup{ACI}}^{(\ell)})}\mathcal{P}_{\textup{CCI}}^{(\ell)}. \label{eq:stat_3}
\end{IEEEeqnarray}

An approximate expression for the BER per subcarrier $i_m$ in the case of $M$-ary QAM \cite{chung2001degrees}, while taking the interference from the PUs into account, is given by
\begin{IEEEeqnarray}{c}
\textup{BER}_{i_m}  \approx   0.2 \: \textup{exp}\left ( -1.6 \frac{\mathcal{P}_{i_m}}{(2^{b_{i_m}} - 1)} \frac{\left | \mathcal{H}_{i_m} \right |^2}{(\sigma^2_n + \mathcal{J}_{i_m})} \right ), \label{eq:BER}
\end{IEEEeqnarray}
where $ \mathcal{H}_{i_m}$ is the channel gain of subcarrier $i_m$ between the SU transmitter and receiver pair and $\sigma^2_n$ is the variance of the additive white Gaussian noise (AWGN).

The multiobjective optimization problem in (\ref{eq:op_1_basic}) can be rewritten as a linear combination of the multiple objectives~as
\setlength{\arraycolsep}{0.0em}
\begin{IEEEeqnarray}{c}
\underset{\mathcal{P}_{i_m} , b_{i_m}}{\textup{Minimize}}  \quad  \mathcal{F}(\mathbf{p}_m,\mathbf{b}_m) = \: \alpha \sum_{i_m = 1}^{N_m}\mathcal{P}_{i_m} - (1-\alpha)\sum_{i_m = 1}^{N_m}b_{i_m}, \label{eq:p1} \nonumber
\end{IEEEeqnarray}
\begin{IEEEeqnarray}{lcl}
\textup{subject to} \quad g_\varrho(\mathbf{p}_m,\mathbf{b}_m) & \: \leq \: &  0, \label{eq:ineq_const}
\end{IEEEeqnarray}
where $\alpha$ ($0 < \alpha < 1$) is a constant which indicates the relative importance of one objective function relative to the other, being selected according to the CR requirements/applications, i.e., minimum power versus maximum throughput, $\varrho = 1, ..., N_m + 2$ is the constraint index, $\mathbf{p}_m = [\mathcal{P}_{1_m},...,\mathcal{P}_{N_m}]^T$ and $\mathbf{b}_m = [b_{1_m},...,b_{N_m}]^T$ are the \textit{$N_m$}-dimensional power and bit distribution vectors, respectively, and
\begin{IEEEeqnarray}{c}
g_\varrho(\mathbf{p}_m,\mathbf{b}_m) = \nonumber \hfill \\ \left\{\begin{matrix}
0.2 \: \sum_{i_m=1}^{N_m} \textup{exp}\left ( \frac{- 1.6 \: \mathcal{C}_{i_m} \mathcal{P}_{i_m}}{2^{b_{i_m}} - 1} \right )  -\: \textup{BER}_{th,{i_m}} \leq 0, \hfill \\ \hfill \varrho = i_m = 1, ..., N_m, \\
\sum_{i_m=1}^{N_m} \mathcal{P}_{i_m} - \Big[\mathcal{P}_{th}, \frac{\nu 10^{0.1L(d_m)}}{-\ln(1 - \Psi_{\textup{CCI}}^{(m)})} \mathcal{P}_{\textup{CCI}}^{(m)}\Big]^- \leq 0, \hfill \\ \hfill \varrho = N_m+1, \\
\sum_{i_m = 1}^{N_m} \mathcal{P}_{i_m} \varpi_{i_m}^{(\ell)} - \frac{\nu}{-\ln(1 - \Psi_{\textup{ACI}}^{(\ell)})}\mathcal{P}_{\textup{CCI}}^{(\ell)} \leq 0, \hfill \varrho = N_m+2,
\end{matrix}\right. \label{eq:ineq}
\end{IEEEeqnarray}
where $\ell = 1, ..., \mathcal{M}$ and $\mathcal{C}_{i_m} = \: \frac{\left | \mathcal{H}_{i_m} \right |^2}{\sigma^2_n + \mathcal{J}_{i_m}}$ is the channel-to-noise-plus-interference ratio for subcarrier $i_m$.

\vspace*{-8pt}
\subsection{Optimization Problem Analysis and Solution}
\vspace*{-5pt}
The optimization problem in (\ref{eq:ineq_const}) can be solved by the method of Lagrange multipliers. Accordingly, the inequality constraints are transformed to equality constraints by adding non-negative slack variables, $\mathcal{Y}^2_\varrho$, $\varrho = 1, ..., N_m+2$ \cite{Boyd2004convex}. Hence, the constraints are given as
\begin{IEEEeqnarray}{c}
\mathcal{G}_\varrho(\mathbf{p}_m,\mathbf{b}_m, \mathbf{y}) = g_\varrho(\mathbf{p}_m,\mathbf{b}_m) + \mathcal{Y}^2_\varrho = 0, \label{eq:slack}
\end{IEEEeqnarray}
where $\mathbf{y} = [\mathcal{Y}^2_1, ..., \mathcal{Y}^{2,(\ell)}_{N_m+2}]^T$ is the vector of slack variables, and the Lagrangian function $\mathcal{L}$ is expressed as
{\small{
\begin{IEEEeqnarray}{lcr}
\mathcal{L}(\mathbf{p}_m,\mathbf{b}_m,\mathbf{y},\boldsymbol\lambda) &{} = {}& \mathcal{F}(\mathbf{p}_m,\mathbf{b}_m) + \sum_{\varrho = 1}^{N_m+2} \lambda_{\varrho} \mathcal{G}_{\varrho}(\mathbf{p}_m,\mathbf{b}_m,\mathbf{y}) \nonumber
\end{IEEEeqnarray}
}}\vspace{-5pt}
{\small{
\begin{IEEEeqnarray}{l}
= \alpha \sum_{i_m = 1}^{N_m}\mathcal{P}_{i_m} - (1-\alpha)\sum_{i_m = 1}^{N_m}b_{i_m} \nonumber \\
+  \sum_{i_m = 1}^{N_m} \lambda_{i_m}\Bigg[ 0.2 \:  \textup{exp}\left ( \frac{- 1.6  \mathcal{C}_{i_m} \mathcal{P}_{i_m}}{2^{b_{i_m}} - 1} \right ) - \textup{BER}_{th,i_m} + \mathcal{Y}_{i_m}^2\Bigg] \nonumber \\
+ \lambda_{N_m+1} \Bigg[\sum_{i_m=1}^{N_m} \mathcal{P}_{i_m} - \Big[\mathcal{P}_{th}, \frac{\nu 10^{0.1L(d_m)}}{-\ln(1 - \Psi_{\textup{CCI}}^{(m)})} \mathcal{P}_{\textup{CCI}}^{(m)}
\Big]^-  + \mathcal{Y}^2_{N_m+1} \Bigg] \nonumber \\
+ \sum_{\ell = 1}^{\mathcal{M}} \lambda_{N_m+2}^{(\ell)} \Bigg[\sum_{i_m = 1}^{N_m} \mathcal{P}_{i_m} \varpi_{i_m}^{(\ell)} - \frac{\nu}{-\ln(1 - \Psi_{\textup{ACI}}^{(\ell)})}\mathcal{P}_{\textup{CCI}}^{(\ell)} + \mathcal{Y}^{2,(\ell)}_{N_m+2} \Bigg],~ \nonumber\\\IEEEeqnarraynumspace
\end{IEEEeqnarray}
}}\hspace{-2pt}where $\boldsymbol\lambda = [\lambda_1, ..., \lambda_{N_m+2}^{(\ell)}]^T$ is the vector of Lagrange multipliers associated with the $N_m+2$ constraints in (\ref{eq:ineq}). A stationary point is found when $\nabla \mathcal{L}(\mathbf{p}_m,\mathbf{b}_m,\mathbf{y},\boldsymbol\lambda) = 0$, which yields
\begin{subequations}
\label{eq:lag}
\begin{IEEEeqnarray}{rCl}
\frac{\partial \mathcal{L}}{\partial \mathcal{P}_{i_m}} &{}  =  {}& \: \alpha - \lambda_{i_m} \frac{(0.2)(1.6) \: \mathcal{C}_{i_m}}{2^{b_{i_m}}-1} \: \textup{exp}\left ( \frac{- 1.6 \: \mathcal{C}_{i_m} \mathcal{P}_{i_m}}{2^{b_{i_m}} - 1} \right ) \nonumber \\ & & \hfill + \lambda_{N_m+1} + \sum_{\ell = 1}^{\mathcal{M}}\varpi_{i_m}^{(\ell)} \lambda_{N_m+2}^{(\ell)} = 0,\label{eq:eq1} \IEEEeqnarraynumspace \\
\frac{\partial \mathcal{L}}{\partial b_{i_m}} &{}  = {}& \: -(1 - \alpha) + \lambda_{i_m} \frac{(0.2)(1.6) (\ln (2)) \: \mathcal{C}_{i_m} \mathcal{P}_{i_m} 2^{b_{i_m}}}{(2^{b_{i_m}}-1)^2} \: \nonumber \\ & & \hfill  \textup{exp}\left ( \frac{- 1.6 \: \mathcal{C}_{i_m} \mathcal{P}_{i_m}}{2^{b_{i_m}} - 1} \right )  = 0,\label{eq:eq2} \IEEEeqnarraynumspace\\
\frac{\partial \mathcal{L}}{\partial \lambda_{i_m}} &{}  = {}& 0.2 \: \textup{exp}\left ( \frac{- 1.6 \: \mathcal{C}_{i_m} \mathcal{P}_{i_m}}{2^{b_{i_m}} - 1} \right ) - \textup{BER}_{th,i_m} + \mathcal{Y}_{i_m}^2 \nonumber \\ & & \hfill = 0, \label{eq:eq3} \IEEEeqnarraynumspace \\
\frac{\partial \mathcal{L}}{\partial \lambda_{N_m+1}} &{}  = {}& \sum_{i_m = 1}^{N_m}\mathcal{P}_{i_m} - \Big[\mathcal{P}_{th}, \frac{\nu 10^{0.1L(d_m)}}{-\ln(1 - \Psi_{\textup{CCI}}^{(m)})} \mathcal{P}_{\textup{CCI}}^{(m)}
\Big]^- \nonumber \\ & & \hfill + \mathcal{Y}_{N_m+1}^2 = 0, \label{eq:eq3_1} \IEEEeqnarraynumspace \\
\frac{\partial \mathcal{L}}{\partial \lambda_{N_m+2}^{(\ell)}} &{}  = {}& \sum_{i_m = 1}^{N_m} \mathcal{P}_{i_m} \varpi_{i_m}^{(\ell)} - \frac{\nu}{-\ln(1 - \Psi_{\textup{ACI}}^{(\ell)})}\mathcal{P}_{\textup{CCI}}^{(\ell)} \nonumber \\ & & \hfill + \mathcal{Y}^{2,(\ell)}_{N_m+2}  = 0, \label{eq:eq3_1new} \IEEEeqnarraynumspace \\
\frac{\partial \mathcal{L}}{\partial \mathcal{Y}_{i,m}} &{}  = {}&  2\lambda_{i_m} \mathcal{Y}_{i_m} = 0, \label{eq:eq4}\\
\frac{\partial \mathcal{L}}{\partial \mathcal{Y}_{N_m+1}} &{}   = {}&  2\lambda_{N_m+1} \: \mathcal{Y}_{N_m+1} = 0, \label{eq:eq4_1} \\
\frac{\partial \mathcal{L}}{\partial \mathcal{Y}_{N_m+2}^{(\ell)}} &{}   = {}& 2\lambda_{N_m+2}^{(\ell)} \: \mathcal{Y}_{N_m+2}^{(\ell)} = 0. \label{eq:eq4_1new}
\end{IEEEeqnarray}
\end{subequations}
It can be seen that (\ref{eq:eq1})-(\ref{eq:eq4_1new}) represent $4N_m + 2\mathcal{M} + 2$ equations in the $4N_m + 2\mathcal{M} + 2$ unknown components of the vectors $\mathbf{p}_m, \mathbf{b}_m, \mathbf{y}$, and $\boldsymbol\lambda$. By solving (\ref{eq:lag}), one obtains the solution $\mathbf{p}^*_m, \mathbf{b}^*_m$. Equation (\ref{eq:eq4}) implies that either $\lambda_{i_m} = 0$ or $\mathcal{Y}_{i_m} = 0$, (\ref{eq:eq4_1}) implies that either $\lambda_{N_m+1} = 0$ or $\mathcal{Y}_{N_m+1} = 0$, and (\ref{eq:eq4_1new}) implies that either $\lambda_{N_m+2}^{(\ell)} = 0$ or $\mathcal{Y}_{N_m+2}^{(\ell)} = 0$. Hence, eight possible cases exist and we are going to investigate each case independently.

--- \textit{Cases 1, 2, 3 , and 4}: In (\ref{eq:lag}), setting $\lambda_{i_m} = 0$ and $\lambda_{N_m+1} = 0$ (case 1)/$\mathcal{Y}_{N_m+1} = 0$ (case 2), or $\lambda_{N_m+2}^{(\ell)} = 0$ (case 3)/$\mathcal{Y}_{N_m+2}^{(\ell)} = 0$ (case 4) results in an underdetermined system, and, hence, no unique solution can be reached.

--- \textit{Case 5}: Setting $\mathcal{Y}_{i_m} = 0$,  $\lambda_{N_m+1} = 0$ (i.e., inactive CCI/total transmit power constraint), and $\lambda_{N_m+2}^{(\ell)} = 0$ (i.e.,  inactive ACI constraint), we can relate $\mathcal{P}_{i_m}$ and $b_{i_m}$ from (\ref{eq:eq1}) and (\ref{eq:eq2}) as
\setlength{\arraycolsep}{0.0em}
\begin{IEEEeqnarray}{c}
\mathcal{P}_{i_m}  =  \frac{1- \alpha}{\alpha \ln(2)}(1 - 2^{-b_{i_m}}), \label{eq:eq8}
\end{IEEEeqnarray}
with $\mathcal{P}_{i_m} \geq 0$ if and only if $b_{i_m} \geq 0$. By substituting (\ref{eq:eq8}) into (\ref{eq:eq3}), one obtains the solution
\vspace*{-2pt}
\begin{IEEEeqnarray}{c}
b_{i_m}^* = \log_2\Bigg[ - \frac{1-\alpha }{\alpha \ln(2)} \frac{1.6 \: \mathcal{C}_{i_m}}{\ln(5 \: \textup{BER}_{th,i_m})}\Bigg]. \label{eq:eq10}
\end{IEEEeqnarray}
Consequently, from (\ref{eq:eq8}) one gets
\setlength{\arraycolsep}{0.0em}
\begin{IEEEeqnarray}{c}
\mathcal{P}_{i_m}^* = \frac{1-\alpha }{\alpha \ln(2)} + \frac{\ln(5 \: \textup{BER}_{th,i_m})}{1.6 \: \mathcal{C}_{i_m}}. \label{eq:eq12}
\end{IEEEeqnarray}
Since we consider $M$-ary QAM, $b_{i_m}$ should be greater than 2. From (\ref{eq:eq10}), to have $b_{i_m} \geq 2$, $\mathcal{C}_{i_m}$, must satisfy the condition
\setlength{\arraycolsep}{0.0em}
\begin{IEEEeqnarray}{c}
\mathcal{C}_{i_m} \geq \mathcal{C}_{th,i_m} = - \frac{4}{1.6} \frac{\alpha \ln(2)}{1-\alpha} \ln(5\textup{BER}_{th,i_m}), i_m = 1, ..., N_m. \nonumber \\ \label{eq:condition}
\end{IEEEeqnarray}

--- \textit{Case 6}: Setting $\mathcal{Y}_{i_m} = 0$,  $\mathcal{Y}_{N_m+1} = 0$ (i.e., active CCI/total transmit power constraint), and  $\lambda_{N_m+2}^{(\ell)} = 0$ (i.e., inactive ACI constraint), similar to \emph{case 5}, we obtain
\begin{IEEEeqnarray}{l}
\mathcal{P}_{i_m}  =  \frac{1- \alpha}{\ln(2)(\alpha + \lambda_{N_m + 1})}(1 - 2^{-b_{i_m}}), \label{eq:eq8new}
\end{IEEEeqnarray}
\begin{IEEEeqnarray}{l}
b_{i_m}^* = \log_2\Bigg[- \frac{1-\alpha }{\ln(2)(\alpha + \lambda_{N_m + 1})} \frac{1.6 \: \mathcal{C}_{i_m}}{\ln(5 \: \textup{BER}_{th,i_m})}\Bigg]. \IEEEeqnarraynumspace \label{eq:eq10new}
\end{IEEEeqnarray}
\begin{IEEEeqnarray}{l}
\mathcal{P}_{i_m}^* = \frac{1-\alpha }{\ln(2)(\alpha + \lambda_{N_m + 1})} + \frac{\ln(5 \: \textup{BER}_{th,i_m})}{1.6 \: \mathcal{C}_{i_m}}, \label{eq:eq12new}
\end{IEEEeqnarray}
where $\lambda_{N_m+1}$ is calculated to satisfy the active CCI/total transmit power constraint in (\ref{eq:eq3_1}). Hence, the value of $\lambda_{N_m+1}$ is found to be
\begin{IEEEeqnarray}{c}
\lambda_{N_m+1} = \hfill \nonumber \\ \frac{\bar{\bar{\mathbb{N}}}_m^a \frac{1 - \alpha}{\ln 2}}{\Big[\mathcal{P}_{th}, \frac{\nu 10^{0.1L(d_m)}}{-\ln(1 - \Psi_{\textup{CCI}}^{(m)})} \mathcal{P}_{\textup{CCI}}^{(m)}\Big]^- - \sum_{i_m \in \mathbb{N}_m^a}^{}\frac{\ln(5 \: \textup{BER}_{th,i_m})}{1.6 \: \mathcal{C}_{i_m}}}  - \alpha, \nonumber \\ \IEEEeqnarraynumspace \label{eq:lambda}
\end{IEEEeqnarray}
where $\bar{\bar{\mathbb{N}}}_m^a$ is the cardinality of the set of active subcarriers $\mathbb{N}_m^a$.

--- \textit{Case 7}: Setting $\mathcal{Y}_{i_m} = 0$, $\lambda_{N_m+1} = 0$ (i.e., inactive CCI/total transmit power constraint), and $\mathcal{Y}_{N_m+2}^{(\ell)} =  0$ (i.e., active ACI constraint), similar to \emph{cases 5 \textup{and} 6}, we obtain
\begin{IEEEeqnarray}{c}
\mathcal{P}_{i_m}  =  \frac{1- \alpha}{\ln(2)(\alpha + \sum_{\ell = 1}^{\mathcal{M}}\varpi_{i_m}^{(\ell)} \lambda_{N_m + 2}^{(\ell)})}(1 - 2^{-b_{i_m}}), \label{eq:eq8new_ACI}
\end{IEEEeqnarray}
\begin{IEEEeqnarray}{c}
b_{i_m}^* = \log_2\Bigg[- \frac{1-\alpha }{\ln(2)(\alpha + \sum_{\ell = 1}^{\mathcal{M}}\varpi_{i_m}^{(\ell)} \lambda_{N_m + 2}^{(\ell)})} \frac{1.6 \mathcal{C}_{i_m}}{\ln(5 \textup{BER}_{th,i_m})}\Bigg]. \nonumber \\ \label{eq:eq10newACI}
\end{IEEEeqnarray}
\begin{IEEEeqnarray}{c}
\mathcal{P}_{i_m}^* = \frac{1-\alpha }{\ln(2)(\alpha + \sum_{\ell = 1}^{\mathcal{M}}\varpi_{i_m}^{(\ell)} \lambda_{N_m + 2}^{(\ell)})} + \frac{\ln(5 \: \textup{BER}_{th,i_m})}{1.6 \: \mathcal{C}_{i_m}}. \label{eq:eq12newACI} \IEEEeqnarraynumspace
\end{IEEEeqnarray}
where $\lambda_{N_m+2}^{(\ell)}$ is calculated numerically using the Newton's method to satisfy the active ACI constraint in (\ref{eq:eq3_1new}).

--- \textit{Case 8}: Setting $\mathcal{Y}_{i_m} = 0$, $\mathcal{Y}_{N_m+1} = 0$ (i.e., active CCI/total transmit power constraint), and  $\mathcal{Y}_{N_m+2}^{(\ell)} =  0$ (i.e., active ACI constraint), similar to the previous \emph{cases}, we obtain
{\small{
\begin{IEEEeqnarray}{c}
\mathcal{P}_{i_m}  = \frac{1- \alpha}{\ln(2)(\alpha + \lambda_{N_m + 1} + \sum_{\ell = 1}^{\mathcal{M}}\varpi_{i_m}^{(\ell)} \lambda_{N_m + 2}^{(\ell)})}(1 - 2^{-b_{i_m}}), \label{eq:eq8new_ACI_CCI} \IEEEeqnarraynumspace
\end{IEEEeqnarray}
}}
{\small{
\begin{IEEEeqnarray}{c}
b_{i_m}^* = \log_2\Bigg[- \frac{1-\alpha }{\ln(2)(\alpha + \lambda_{N_m + 1} + \sum_{\ell = 1}^{\mathcal{M}}\varpi_{i_m}^{(\ell)} \lambda_{N_m + 2}^{(\ell)})} \nonumber \\ \hspace{4.5cm} \frac{1.6 \mathcal{C}_{i_m}}{\ln(5 \textup{BER}_{th,i_m})}\Bigg], \label{eq:eq10newACI_CCI}
\end{IEEEeqnarray}
}}
{\small{
\begin{IEEEeqnarray}{c}
\mathcal{P}_{i_m}^* = \frac{1-\alpha }{\ln(2)(\alpha + \lambda_{N_m + 1} +  \sum_{\ell = 1}^{\mathcal{M}}\varpi_{i_m}^{(\ell)} \lambda_{N_m + 2}^{(\ell)})} + \frac{\ln(5 \: \textup{BER}_{th,i_m})}{1.6 \: \mathcal{C}_{i_m}}, \label{eq:eq12newACI_CCI} \nonumber \\
\end{IEEEeqnarray}
}}\hspace{-1pt}where $\lambda_{N_m+1}$ and $\lambda_{N_m+2}^{(\ell)}$ are calculated numerically to satisfy the active CCI/total transmit power and ACI constraints in (\ref{eq:eq3_1}) and (\ref{eq:eq3_1new}), respectively.

\vspace{-1pt}
The obtained solution ($\mathbf{p}_m^*,\mathbf{b}_m^*$) represents a minimum of $\mathcal{F}(\mathbf{p}_m,\mathbf{b}_m)$ as the Karush-Kuhn-Tucker (KKT) conditions \cite{Boyd2004convex} are satisfied; the proof is not included due to space limitations. Please note that the optimization problem in (\ref{eq:ineq_const}) is not convex, and the obtained solution is not guaranteed to be a global optimum. In the next section, we compare the local optimum results to the global optimum results achieved through an exhaustive search to 1) characterize the gap to the global optimum solution and 2) characterize the gap to the equivalent discrete optimization problem (i.e., with integer constraints on $b_{i_m}$).
\vspace*{-10pt}

\subsection{Proposed Joint Bit and Power Loading Algorithm}
The proposed algorithm can be formally stated as follows \vspace*{-6pt}
\floatname{algorithm}{}
\begin{algorithm}
\renewcommand{\thealgorithm}{}
\caption{\textbf{Proposed Algorithm}}
\begin{algorithmic}[1]
\small
\State \textbf{INPUT} $\sigma^2_n$, $\mathcal{H}_{i_m}$, $\textup{BER}_{th,i_m}$, $\alpha$, $\mathcal{P}_{th}$, $\mathcal{P}_{\textup{CCI}}^{(m)}$, $\mathcal{P}_{\textup{CCI}}^{(\ell)}$, $\nu$, $\Psi_{\textup{CCI}}^{(m)}$, $\Psi_{\textup{ACI}}^{(\ell)}$, and PUs information.
\For{$i_m$ = 1, ..., $N_m$}
\If{$\mathcal{C}_{i_m} \geq \mathcal{C}_{th,i_m} =  - \frac{4}{1.6} \: \frac{\alpha \ln(2)}{1-\alpha} \: \ln(5\:\textup{BER}_{th,i_m})$}
\State  - $b_{i_m}^*$ and $\mathcal{P}_{i_m}^*$ are given by (\ref{eq:eq10}) and (\ref{eq:eq12}), respectively.
\Else
\State Null the corresponding subcarrier $i_m$.
\EndIf
\EndFor
\If {$\sum_{i_m=1}^{N_m} \mathcal{P}_{i_m} \geq \Big[\mathcal{P}_{th}, \frac{\nu 10^{0.1L(d_m)}}{-\ln(1 - \Psi_{\textup{CCI}}^{(m)})} \mathcal{P}_{\textup{CCI}}^{(m)}\Big]^-$ and $\sum_{i_m = 1}^{N_m} \mathcal{P}_{i_m} \varpi_{i_m}^{(\ell)} \leq \frac{\nu}{-\ln(1 - \Psi_{\textup{ACI}}^{(\ell)})}\mathcal{P}_{\textup{CCI}}^{(\ell)}$}
\State  - $b_{i_m}^*$ and $\mathcal{P}_{i_m}^*$ are given by (\ref{eq:eq10new}) and (\ref{eq:eq12new}), respectively.
\State - $\lambda_{N_m + 1}$ is given by (\ref{eq:lambda}) and $\lambda_{N_m + 2}^{(\ell)} = 0$.
\ElsIf {$\sum_{i_m=1}^{N_m} \mathcal{P}_{i_m} \leq  \Big[\mathcal{P}_{th}, \frac{\nu 10^{0.1L(d_m)}}{-\ln(1 - \Psi_{\textup{CCI}}^{(m)})} \mathcal{P}_{\textup{CCI}}^{(m)}\Big]^-$ and $\sum_{i_m = 1}^{N_m} \mathcal{P}_{i_m} \varpi_{i_m}^{(\ell)} \geq \frac{\nu}{-\ln(1 - \Psi_{\textup{ACI}}^{(\ell)})}\mathcal{P}_{\textup{CCI}}^{(\ell)}$}
\State  - $b_{i_m}^*$ and $\mathcal{P}_{i_m}^*$ are given by (\ref{eq:eq10newACI}) and (\ref{eq:eq12newACI}), respectively.
\State - $\lambda_{N_m + 1} = 0$ and $\lambda_{N_m + 2}^{(\ell)}$ are calculated to satisfy $\sum_{i_m = 1}^{N_m} \mathcal{P}_{i_m} \varpi_{i_m}^{(\ell)} = \frac{\nu}{-\ln(1 - \Psi_{\textup{ACI}}^{(\ell)})}\mathcal{P}_{\textup{CCI}}^{(\ell)}$
\Else \: \textbf{if} {\: $\sum_{i_m=1}^{N_m} \mathcal{P}_{i_m} \geq \Big[\mathcal{P}_{th}, \frac{\nu 10^{0.1L(d_m)}}{-\ln(1 - \Psi_{\textup{CCI}}^{(m)})} \mathcal{P}_{\textup{CCI}}^{(m)}
\Big]^-$ and $\sum_{i_m = 1}^{N_m} \mathcal{P}_{i_m} \varpi_{i_m}^{(\ell)} \geq \frac{\nu}{-\ln(1 - \Psi_{\textup{ACI}}^{(\ell)})}\mathcal{P}_{\textup{CCI}}^{(\ell)}$}
\State  - $b_{i_m}^*$ and $\mathcal{P}_{i_m}^*$ are given by (\ref{eq:eq10newACI_CCI}) and (\ref{eq:eq12newACI_CCI}), respectively.
\State - $\lambda_{N_m + 1}$ and $\lambda_{N_m + 2}^{(\ell)}$ are calculated to satisfy $\sum_{i_m=1}^{N_m} \mathcal{P}_{i_m} = \Big[\mathcal{P}_{th}, \frac{\nu 10^{0.1L(d_m)}}{-\ln(1 - \Psi_{\textup{CCI}}^{(m)})} \mathcal{P}_{\textup{CCI}}^{(m)}\Big]^-$ and $\sum_{i_m = 1}^{N_m} \mathcal{P}_{i_m} \varpi_{i_m}^{(\ell)} = \frac{\nu}{-\ln(1 - \Psi_{\textup{ACI}}^{(\ell)})}\mathcal{P}_{\textup{CCI}}^{(\ell)}$, respectively.
\EndIf
\State - $b^*_{i_m,final}$ $\leftarrow$ Round $b_{i_m}^*$ to the nearest integer.
%
\State - $\mathcal{P}^*_{i_m,final}$ $\leftarrow$ Recalculate $\mathcal{P}_{i_m}^*$ according to (\ref{eq:BER}).
\State - If the conditions on the CCI/total transmit power and the ACI are violated due to rounding, decrement the number of bits on the subcarrier that has the largest $\Delta \mathcal{P}_{i_m}(b_{i_m}) = \mathcal{P}_{i_m}(b_{i_m}) - \mathcal{P}_{i_m}(b_{i_m} - 1)$ until satisfied.
\State \textbf{OUTPUT} $b^*_{i_m,final}$ and $\mathcal{P}^*_{i_m,final}$, $i_m$ = 1, ..., $N_m$.
\end{algorithmic}
\end{algorithm}

\vspace*{-15pt}
\section{Numerical Results}
\vspace*{-4pt}
In this section, we present illustrative numerical results for the proposed allocation algorithm. Without loss of generality,
we assume that the OFDM SU coexists with one adjacent PU and one co-channel PU. The OFDM SU transmission parameters are as follows: number of subcarriers $N_m = 128$, symbol duration $T_{s,m} = 102.4 \: \mu \textup{sec}$, and subcarrier spacing $\Delta f_m = 9.7656$ kHz. The path loss parameters are as follows: exponent $= 4$, wavelength $= 0.33\:\textup{meters}$, distance to the $\ell$th PU receiver  $d_{\ell} = 1$ km, distance to the $m$th PU receiver $d_m = 5$ km, and reference distance $d_0 = 500$ m. $\textup{BER}_{th,i_m}$ is assumed to be the same for all subcarriers and set to $10^{-4}$. 
$\sigma_n^2$ is assumed to be $10^{-3} \mu$W and the PUs signals are assumed to be elliptically-filtered white random processes \cite{hasan2009energy, zhao2010power, bansal2008optimal, bansal2011adaptive}.
Representative results are presented in this section, which were obtained through Monte Carlo trials for $10^{4}$ channel realizations. 
Unless otherwise mentioned, $\alpha = 0.5$ and $\Psi_{\textup{CCI}} = \Psi_{\textup{ACI}} = 0.9$.

\begin{figure}[!t]
	\centering
		\includegraphics[width=0.50\textwidth]{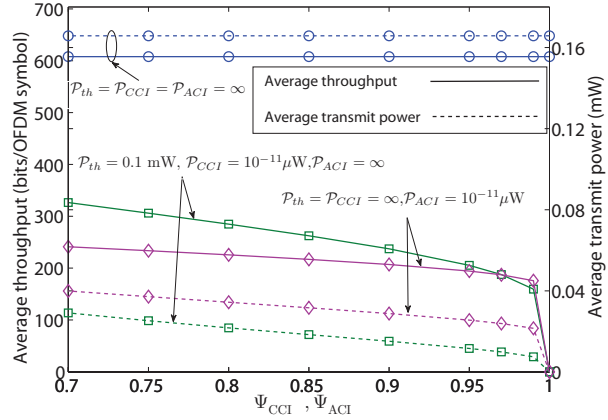}
	\caption{Effect of $\Psi_{\textup{CCI}}$ and $\Psi_{\textup{ACI}}$ on the SU performance for different values of $\mathcal{P}_{th}$, $\mathcal{P}_{\textup{CCI}}$, and $\mathcal{P}_{\textup{ACI}}$.}
	\label{fig:performance_gamma}
\end{figure}

In Fig. \ref{fig:performance_gamma}, the average throughput and transmit power are plotted as a function of the probabilities $\Psi_{\textup{CCI}}$ and $\Psi_{\textup{ACI}}$, for different values of $\mathcal{P}_{th}$, $\mathcal{P}_{\textup{CCI}}$, and $\mathcal{P}_{\textup{ACI}}$.  As expected, for $\mathcal{P}_{th}= \mathcal{P}_{\textup{CCI}} = \infty$ and $\mathcal{P}_{\textup{ACI}} = \infty$, increasing the value of  $\Psi_{\textup{CCI}}$ and $\Psi_{\textup{ACI}}$ has no effect on the achieved average throughput and transmit power, as the CCI and ACI constraints are inactive.
For other values of $\mathcal{P}_{th}$, $\mathcal{P}_{\textup{CCI}}$, and $\mathcal{P}_{\textup{ACI}}$, increasing the value of the probabilities $\Psi_{\textup{CCI}}$ and $\Psi_{\textup{ACI}}$, slightly decreases the achieved average throughput and transmit power in order to meet such tight statistical constraints (i.e., meeting the CCI and ACI constraints with higher probabilities). The achieved average throughput and transmit power drop to zero for $\Psi_{\textup{CCI}} = \Psi_{\textup{ACI}} = 1$ as the proposed algorithm cannot meet such stringent requirements of satisfying the active CCI and the ACI constraints all the time, without knowledge of the instantaneous channel gains.


\begin{figure}[!t]
	\centering
		\includegraphics[width=0.50\textwidth]{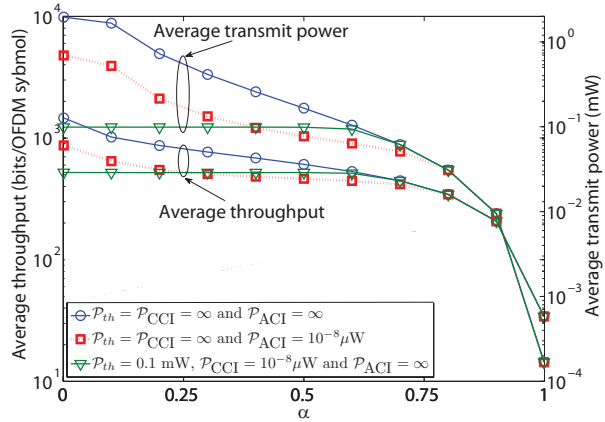}
	\caption{Effect of $\alpha$ on the SU performance for different values of $\mathcal{P}_{th}$, $\mathcal{P}_{\textup{CCI}}$, and $\mathcal{P}_{\textup{ACI}}$.}
	\label{fig:proposed_alpha}
\end{figure}

Fig. \ref{fig:proposed_alpha} shows the average throughput and transmit power as a function of the weighting factor $\alpha$, for different values of $\mathcal{P}_{th}, \mathcal{P}_{\textup{CCI}}$, and $\mathcal{P}_{\textup{ACI}}$. For $\mathcal{P}_{th} = \mathcal{P}_{\textup{CCI}} = \infty$ and $\mathcal{P}_{\textup{ACI}} = \infty$, one can notice that an increase of the weighting factor $\alpha$ yields a decrease of both the average throughput and  transmit power. This can be explained as follows: by increasing $\alpha$, more weight is given to the transmit power  minimization (the minimum transmit power is further reduced), whereas less weight is given to the throughput maximization (the maximum throughput is reduced), according to the problem formulation. 
Similar behaviour is noticed for $\mathcal{P}_{th} = \mathcal{P}_{\textup{CCI}} = \infty$ and $\mathcal{P}_{\textup{ACI}} = 10^{-8} \mu$W with reduced values of the average throughput and transmit power for lower values of $\alpha$ due to the active ACI constraint.
For $\mathcal{P}_{th} = 0.1\:$mW and $\mathcal{P}_{\textup{CCI}} = 10^{-8}\mu$W  and $\mathcal{P}_{\textup{ACI}} = \infty$, the average throughput and transmit power are similar to their respective values if the total transmit power is less than $\big[\mathcal{P}_{th}, \frac{\nu 10^{0.1L(d_m)}}{-\ln(1 - \Psi_{\textup{CCI}})} \mathcal{P}_{\textup{CCI}}\big]^- = \big[0.1 \: \textup{mW}, 15.4307 \; \textup{mW}\big]^- = 0.1 \: \textup{mW}$, while they saturate if the total transmit power exceeds 0.1 mW. 
Fig. \ref{fig:proposed_alpha} illustrates the benefit of introducing such a weighting factor in our problem formulation to tune the average throughput and transmit power levels as needed by the CR system.

\begin{figure}[!t]
	\centering
		\includegraphics[width=0.480\textwidth]{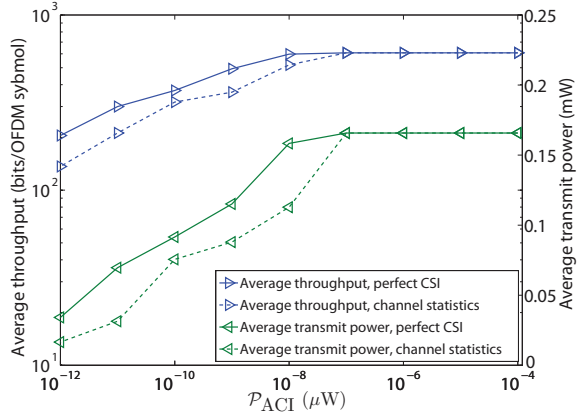}
	\caption{Effect of $\mathcal{P}_{\textup{ACI}}$ on the SU performance for $\mathcal{P}_{th} = \mathcal{P}_{\textup{CCI}} = \infty$, with perfect CSI and channel statistics, respectively.}
	\label{fig:performance_ACI}
\end{figure}

Fig. \ref{fig:performance_ACI} depicts the average throughput and transmit power as a function of the ACI threshold $\mathcal{P}_{\textup{ACI}}$, for $\mathcal{P}_{th} = \mathcal{P}_{\textup{CCI}} = \infty$ and with knowledge of the perfect CSI and channel statistics, respectively. As can be seen for both cases of channel knowledge, the average throughput and transmit power increase as $\mathcal{P}_{\textup{ACI}}$ increases, and saturate for higher values of $\mathcal{P}_{\textup{ACI}}$. This behaviour can be explained, as for lower values of $\mathcal{P}_{\textup{ACI}}$ the ACI constraint is active and it affects the total transmit power. Increasing $\mathcal{P}_{\textup{ACI}}$ results in a corresponding increase in both the average throughput and total transmit power. For higher values of $\mathcal{P}_{\textup{ACI}}$, the ACI constraint is inactive and the achieved throughput and transmit power saturate. As expected, the same performance is achieved for both perfect CSI and channel statistics knowledge for higher values of  $\mathcal{P}_{\textup{ACI}}$; this is because the ACI constraint is inactive, i.e., $\mathcal{P}_{\textup{ACI}} = \infty$ (please note that the CCI is inactive), and it will not be violated regardless of the channel knowledge. On the other hand, for lower values of $\mathcal{P}_{\textup{ACI}}$ and with only knowledge of the channel statistics, the achieved average throughput and transmit power degrade when compared to the case of perfect CSI.

\begin{figure}[!t]
	\centering
		\includegraphics[width=0.50\textwidth]{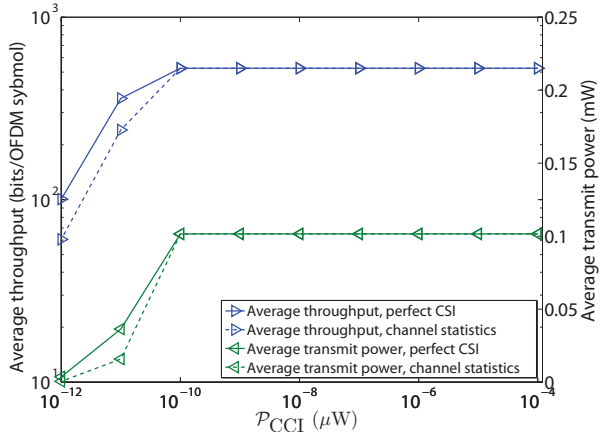}
	\caption{Effect of $\mathcal{P}_{\textup{CCI}}$ on the SU performance for $\mathcal{P}_{th} = 0.1$ mW and $\mathcal{P}_{\textup{ACI}} = \infty$, with perfect CSI and channel statistics, respectively.}
	\label{fig:performance_CCI}
\end{figure}

In Fig. \ref{fig:performance_CCI}, we plot the average throughput and transmit power as a function of the CCI threshold $\mathcal{P}_{\textup{CCI}}$, for $\mathcal{P}_{th} = 0.1$ mW and $\mathcal{P}_{\textup{ACI}} = \infty$, and with knowledge of the perfect CSI and channel statistics, respectively. As can be seen for both cases of channel knowledge, the average throughput and transmit power increase as $\mathcal{P}_{\textup{CCI}}$ increases, and saturate for higher values of $\mathcal{P}_{\textup{CCI}}$. This can be explained, as for lower values of $\mathcal{P}_{\textup{CCI}}$, $\big[0.1 \: \textup{mW}, \frac{\nu 10^{0.1L(d_m)}}{-\ln(1 - \Psi_{\textup{CCI}})} \mathcal{P}_{\textup{CCI}}\big]^- = \frac{\nu 10^{0.1L(d_m)}}{-\ln(1 - \Psi_{\textup{CCI}})} \mathcal{P}_{\textup{CCI}}$. Hence, the CCI constraint is active and affects the total transmit power. Increasing $\mathcal{P}_{\textup{CCI}}$ results in a corresponding increase in both the average throughput and transmit power. For higher values of $\mathcal{P}_{\textup{CCI}}$, $\big[0.1 \: \textup{mW}, \frac{\nu 10^{0.1L(d_m)}}{-\ln(1 - \Psi_{\textup{CCI}})} \mathcal{P}_{\textup{CCI}}\big]^- =  0.1 \: \textup{mW}$ and the transmit power is limited by the value of $\mathcal{P}_{th} = 0.1$ mW, while the achieved throughput saturates accordingly. Similar to the discussion in Fig. \ref{fig:performance_ACI}, the performance degrades for the case when only the channel statistics are known if the CCI constraint is active, i.e., at lower values of $\mathcal{P}_{\textup{CCI}}$. On the other hand, the same performance is achieved for both cases of the channel knowledge for higher values of $\mathcal{P}_{\textup{CCI}}$ (please note that the ACI constraint is inactive).

\begin{figure}[!t]
	\centering
		\includegraphics[width=0.50\textwidth]{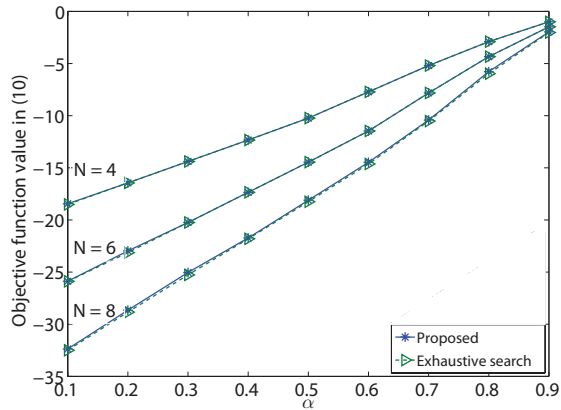}
	\caption{Objective function for the proposed algorithm and the exhaustive search for $\mathcal{P}_{th}$ = 5 $\mu$W, $\mathcal{P}_{\textup{CCI}} = \mathcal{P}_{\textup{ACI}} = 10^{-10} \mu$W.}
	\label{fig:ex}
\end{figure}

Fig. \ref{fig:ex} compares the objective function achieved with the proposed algorithm and an exhaustive search that finds the discretized global optimal allocation for the problem in (\ref{eq:ineq_const}) for $\mathcal{P}_{th}$ = 5 $\mu$W, $\mathcal{P}_{\textup{CCI}} = \mathcal{P}_{\textup{ACI}} = 10^{-10} \mu$W. Results are presented for  a small number of subcarriers $N_m$ = 4, 6, and 8, such that the exhaustive search is feasible. The exhaustive search tests all  possible combinations of the bit and power allocations (the power per subcarrier is calculated from the discrete value of the bit allocation and $\textup{BER}_{th,i}$) and selects the pair with the least objective function value. As one can notice, the proposed algorithm approaches the optimal results of the exhaustive search.  The computational complexity of the proposed algorithm is of $\mathcal{O}(N^2)$ (the complexity analysis is not provided due to the space limitations), which is significantly lower than $\mathcal{O}(N!)$ of the exhaustive search.

\vspace*{-8pt}
\section{Conclusions}
\vspace*{-4pt}
In this paper, we proposed a joint bit and power loading algorithm that maximizes the OFDM SU throughput and minimizes its transmit power while guaranteeing a target BER and a  total transmit power threshold for the SU, and ensuring that the CCI and ACI are below certain thresholds with predefined probabilities. Unlike most of the work in the literature, the proposed algorithm does not require instantaneous channel information feedback between the SU transmitter and the PUs receivers. Closed-form expressions were derived for the close-to-optimal bit and power distributions.
Simulation results showed the flexibility of the proposed algorithm to tune for various power and throughput levels as needed by the CR system while meeting the constraints, with low computational complexity.
\vspace*{-15pt} 

\end{document}